\documentclass[a4paper,11pt]{article}
\usepackage[english]{babel}
\usepackage{amsmath,amsfonts,color,amssymb}
\usepackage{epsfig}
\usepackage{color}

\newcommand{\be}{\begin{equation}}
\newcommand{\ee}{\end{equation}}
\newcommand{\bea}{\begin{eqnarray}}
\newcommand{\eea}{\end{eqnarray}}

\renewcommand{\d}{{\mathbf d}}
\newcommand{\uno}{{\mathbf 1}}
\newcommand{\T}{{\mathbf T}}

\newcommand{\defin}{\mathop{=}^{\text{def}}}

\makeatletter
\renewcommand{\@makecaption}[2]{
   \vskip\abovecaptionskip
   \sbox\@tempboxa{#1. #2}%
   \ifdim \wd\@tempboxa >\hsize
     #1. #2\par
   \else
     \global \@minipagefalse
     \hb@xt@\hsize{\hfil\box\@tempboxa\hfil}%
   \fi
   \vskip\belowcaptionskip}
\makeatother

\numberwithin{equation}{section}
\title{ON THE INTEGRABLE STRUCTURE OF THE ISING MODEL}
\author{Alessandro Nigro\\Dipartimento di Fisica and INFN- Sezione di Milano\\
 Universit\`a degli Studi di Milano I\\
 Via Celoria 16, I-20133 Milano, Italy\\
 Alessandro.Nigro@mi.infn.it}

\begin{document}
\maketitle
\begin{abstract} Starting from the lattice $A_3$ realization of the Ising model defined on a strip with integrable boundary conditions, the exact 
spectrum (including excited states) of  all the local integrals of motion is derived in the continuum limit by means of TBA techniques. It is also possible to follow the massive flow of this spectrum between the UV $c=1/2$ conformal fixed point and the massive IR theory. The UV expression of the eigenstates of such integrals of motion in terms of Virasoro modes is found to have only rational coefficients and their fermionic representation turns out to be simply related to the quantum numbers describing the spectrum. \end{abstract}

\bigskip 

\section*{Introduction}
It is well known that a deep connection exists between integrable models of statistical mechanics and integrable quantum field theories. In particular in quantum field theory the Yang Baxter equation (YBE) plays an important role as a constraint on the 2-particles $S-$matrix. On the other hand in statistical mechanics the same equation appears as an equation satisfied by the Boltzmann Weights. Boltzmann weights satisfying the YBE are then used to build families of commuting transfer matrices, which is another quite general feature of integrability for lattice models.\\
If we take as a prototype the $\mathbf{A}_n$ RSOS models one finds that the Boltzmann Weights can be understood as elliptic solutions of the YBE, and actually it is possible to recognize that the solutions one finds are connected with $S-$matrices in the Sine-Gordon theory \cite{shadowworld}.\\
Furthermore it is well known that upon a suitable restriction of the couplings the Sine-Gordon model is equivalent to minimal conformal field theories \cite{residual}\cite{minimal}\cite{hidden}, and actually also the $\mathbf{A}_n$ RSOS have been shown to be in the universality class of minimal CFTs \cite{pn}.\\
Nonetheless  the $\mathbf{A}_n$ are in correspondence with CFTs only in an appropriate continuum limit called UV scaling limit. In general there will be a continuum scaling limit depending on a mass parameter $\mu$ which will generate a RG flow to a massive IR theory, where the relevant processes will essentially be the scattering of kinks. \\
The continuum field theory corresponding to $\mathbf{A}_n$ models can be interpreted as a $\phi_{1,3}$ thermal perturbation of the $\mathcal{M}_{n,n+1}$ minimal conformal field theory \cite{hidden} . Such a perturbation is known to be integrable \cite{zam}\cite{goshzam}, this means that there exists an infinite number of commuting currents which remain conserved in the perturbed theory.\\
In particular the first conserved quantity is the energy, if one considers its value on the vacuum state it is well known that this is proportional to the central charge of the underlying CFT in the UV limit. The flow to the IR of such a quantity represents an example of the famous $c-$theorem.\\
A powerful tool for having access to the vacuum energy is the Thermodynamic Bethe Ansatz (TBA)(see for example \cite{mussardo}), which in some cases has been generalized to excited states \cite{pn}\cite{marcio}\cite{doreytateo}\cite{fendley}. \\
In this work we derive the excited TBA equations for the $\mathbf{A}_3$ model on a strip with integrable boundary conditions by diagonalizing the transfer matrix. We then proceed to define the continuum scaling limit of the transfer matrix eigenvalues which we then use as generating functions for some quantities which we eventually identify with the conserved quantities of the thermally perturbed conformal field theory.\\
It is then possible to analytically follow all the conserved quantities along the massive flow to the IR theory. Comparison of the results which are obtained in the UV limit with the spectrum of the BLZ local integrals of motion provides an exact identification of the conserved quantities and allows to put the lattice boundary conditions in correspondence with the CFT operator content of the theory.\\
The eigenstates of the BLZ integrals of motion are computed in their Virasoro form, and once expressed in terms of the fermion field turn out to be labelled by the same quantum numbers which label the exact formula for their eigenvalues which has been derived independently through TBA.

\section{The $A_3$ Model}
The $\mathbf{A}_3$ model is a lattice model which provides a convenient realization of the Ising model. It is built on a square lattice where to each site $j$ is assigned a height variable $a_j\in 
\{1,2,3\}$. The local height variables $a_j$ are constrained to satisfy an adjacency rule which holds for $i,j$ nearest 
neighbours:
\be |a_i-a_j|=1 \ee
The model is characterized by its Boltzmann Weights \cite{bp}, which can be of two different types: bulk weights and boundary 
weights.
The only nonvanishing bulk weights are:
\begin{equation}\nonumber W\left(\begin{array}{cc} a\pm 1  & a \\ a & a\mp 1 \end{array} 
\right|u\Bigg)=\frac{\theta_1(\lambda-u,q)}{\theta_1(\lambda,q)} \end{equation}
\begin{equation}\label{boltzwe}W\left(\begin{array}{cc} a  & a\pm 1 \\ a\mp 1 & a \end{array} 
\right|u\Bigg)=\frac{\sqrt{\theta_1((a-1)\lambda,q)\theta_1((a+1)\lambda,q)}}{\theta_1(a\lambda,q)}\frac{\theta_1(u,q)}{\theta_1( 
\lambda,q)} \end{equation}
\begin{equation}\nonumber W\left(\begin{array}{cc} a  & a\pm 1 \\ a\pm 1 & a \end{array} \right|u\Bigg)=\frac{\theta_1(a\lambda\pm 
u,q)}{\theta_1(a\lambda,q))} \end{equation}
while the non vanishing boundary weights are:
\begin{equation}\label{boundwe}\begin{split}K_L\left(\left.\begin{matrix}a&\\[-2mm]&a\mp 1\\[-2mm]a&\end{matrix}
\right| \,u \right)= \sqrt{\frac{\theta_1((a\mp 1)\lambda,q)}{\theta_1(a\lambda,q)}}\frac{\theta_4(u\mp\xi_L(a),q)\theta_4(u\pm 
a\lambda\pm\xi_L(a),q)}{\theta_4^2(\lambda,q)}\end{split} \end{equation}
\begin{equation}\begin{split}K_R\left(\left.\begin{matrix}&a\\[-2mm]a\mp 1&\\[-2mm]&a\end{matrix}
\right| \,u \right)= \sqrt{\frac{\theta_1((a\mp 1)\lambda,q)}{\theta_1(a\lambda,q)}}\frac{\theta_4(u\mp\xi_R(a),q)\theta_4(u\pm 
a\lambda\pm\xi_R(a),q)}{\theta_4^2(\lambda,q)}\end{split} \end{equation}
where $\lambda=\pi/4$ is the so called crossing parameter, $u$ is the spectral parameter and the $\xi$ are related to the choice 
of boundary condition.\\
We also define the elliptic theta functions of nome $q$ as:
\be \left\{ \begin{array}{l}\theta_1(u,q)=2q^{1/4}\displaystyle\sum_{k=0}^{\infty}(-1)^k q^{k(k+1)}\sin((2k+1)u) \quad |q|<1 \\ 
\theta_2(u,q)=2q^{1/4}\displaystyle\sum_{k=0}^{\infty} q^{k(k+1)}\cos((2k+1)u) \quad |q|<1 \\ \theta_4(u,q)=1+2\displaystyle\sum_{k=1}^\infty(-1)^{k}q^{k^2}\cos(2k u) \quad |q|<1 \end{array} \right. \ee 
this is the so called $q$-series, which will prove more useful to our goal, more typical definitions of these functions are given 
in terms of infinite products.\\
It is important to remark that the role of the nome $q$ is to control the criticality of the model, which becomes critical as 
$q\to 0$. In what will follow we will focus on the region $0<q<1$ which is the so called regime III  of \cite{pn}, actually this regime corresponds to a low temperature phase, but because of  the duality symmetry of the Ising model this is the same as a high temperature phase.\\
Now, in terms of the above objects it is known that the model admits a transfer matrix description (on a lattice of width N):
\begin{equation}\label{tmat}\begin{split} &\big<a_1\ldots a_{N+1}\big|\mathbf{T}(u)\big|b_1\ldots b_{N+1}\big>=\sum_{c_1\ldots 
c_{N+1}}K_L\left(\left.\begin{matrix}b_{1}&\\[-2mm]&c_{1}\\[-2mm]a_{1}&\end{matrix}
\right| \,\lambda-u \right)\cdot\\
&\cdot\Bigg[\prod_{j=1}^N W\left(\begin{array}{cc} c_j  & c_{j+1} \\ a_j & a_{j+1} \end{array} 
\right|u\Bigg)W\left(\begin{array}{cc} b_j  & b_{j+1} \\ c_j & c_{j+1} \end{array} 
\right|\lambda-u\Bigg)\Bigg]K_R\left(\left.\begin{matrix}&b_{N+1}\\[-2mm]c_{N+1}&\\[-2mm]&a_{N+1}\end{matrix}
\right| \,u \right)  \end{split}  \end{equation}
such a transfer matrices form a one parameter commuting family with respect to the spectral parameter $u$, and it is well known 
that this property makes the model integrable.\\
The transfer matrix $\T(u)$ satisfies the following functional equation:
 \setlength{\arraycolsep}{0.5mm}
\bea \label{eqfun}
\T(u)\T(u+\lambda) &
=&  \Big( \uno +\d(u) \Big) 
~ F_N(u) ~S(u,\xi_L,\xi_R)=\mathcal{F}(u,q)  \\[2mm]
 \nonumber 
\eea
\setlength{\arraycolsep}{1.1mm}
With
\be F_N(u,q)=\Bigg[ \frac{\theta_1(u-\lambda)\theta_1(u+\lambda)}{\theta_1(\lambda)^2} \Bigg]^{2N} \ee
\be S(u,\xi_L,\xi_R)=\frac{\theta_1(2u-2\lambda)\theta_1(2u+2\lambda)}
{\theta_1(2u-\lambda)\theta_1(2u+\lambda)} ~ \mathcal{A}_L(u,q,\xi_L,a_L)\mathcal{A}_R(u,q,\xi_R,a_R) \ee
where $\d$ is a matrix proportional to the identity that takes the form:
\be \d(u,q)=\uno(-1)^N \Bigg[ \frac{\theta_1(u)\theta_1(u-2\lambda)}
{\theta_1(u-\lambda)\theta_1(u+\lambda)} \Bigg]^{2N}\Big\{\frac{\theta_1(2u)^2}
{\theta_1(2u-2\lambda)\theta_1(2u+2\lambda)}\Big\} \Big(\mathcal{B}_L(u,\xi_L,a_L)\mathcal{B}_R(u,\xi_R,a_R)\Big)\ee
being
\be 
\mathcal{B}_L=e^{i\pi a_L}\frac{\theta_4(u+\pi/4-\xi_L)\theta_4(u-\pi/4-\xi_L)\theta_4(u+\pi/4+\xi_L)\theta_4(u-\pi/4+\xi_L)}{\theta_4(u-a_L\pi/4-\xi_L)
\theta_4(u+a_L\pi/4+\xi_L) \theta_4(u-(a_L+2)\pi/4-\xi_L)\theta_4(u+(a_L-2)\pi/4+\xi_L)} \ee
\be 
\mathcal{B}_R=e^{-i\pi a_R}\frac{\theta_4(u+\pi/4-\xi_R)\theta_4(u-\pi/4-\xi_R)\theta_4(u+\pi/4+\xi_R)\theta_4(u-\pi/4+\xi_R)}{\theta_4(u-a_R\pi/4-\xi_R)
\theta_4(u+a_R\pi/4+\xi_R) \theta_4(u-(a_R+2)\pi/4-\xi_R)\theta_4(u+(a_R-2)\pi/4+\xi_R)} \ee
\be\mathcal{A}_L(u,q,\xi_L,a_L)=\frac{\theta_4(u-\xi_L)\theta_4(u+\xi_L)\theta_4(u+a_L\pi/4+\xi_L)\theta_4(u-a_L\pi/4-\xi_L)}{\theta_4(\lambda)^4} \ee
\be\mathcal{A}_R(u,q,\xi_R,a_R)=\frac{\theta_4(u-\xi_R)\theta_4(u+\xi_R)\theta_4(u+a_R\pi/4+\xi_R)\theta_4(u-a_R\pi/4-\xi_R)}{\theta_4(\lambda)^4} \ee
The phases in the $\mathcal{B}$ terms may seem strange, but they turn out to be necessary. This fact has been observed also in \cite{unpublished} where the TBA  equations for the $\mathbf{A}_3$ model were derived.  \\
Such a matrix satisfies a functional equation which for obvious reasons is called the Inversion Equation:
\be
\d(u)\d(u+\lambda)=\uno
\ee 
As a consequence of the simple form of the $\d$ matrix (which for more complicated models is not diagonal but is expressed in 
terms of $\T$ itself), we have that the Functional equation written in terms of the eigenvalues $T$ of $\T$ is independent of the 
eigenvalue under consideration.\\
Before moving on to discuss the TBA equations, it is useful to spend some word to comment on the periodicities of the transfer 
matrix $\T(u)$. Such periodicities come directly from the properties of the elliptic $\theta$ functions and read:
\be \T(u+\pi)=\T(u) \ee
\be \T(u-i\log q)=\T(u)\ee
As a consequence we have that $\T$ is a doubly periodic function which is completely defined by its analytic properties inside a 
rectangle that we may take as:
\be \label{cell} (-\frac{\lambda}{2},\frac{7}{2}\lambda)\times i(\frac{1}{2}\log(q),-\frac{1}{2}\log(q)) \ee
If we now consider the functional equation (\ref{eqfun}) it is clear that $D$'s periodicities are inherited by the righthand side 
$\mathcal{F}(u)$ so that the object of our interest will be the zeroes of $\mathcal{F}$ inside the periodicity rectangle 
(\ref{cell}).\\
Such zeroes can be shown by numerical analisys (and indeed analytically in the critical limit \cite{ret1} ) to be organized on 
lines parallel to $u=\lambda/2+ix$ with periodicity $\lambda$. It can be argued (see again \cite{ret1} for the critical case) that 
as a consequence of the periodicities and of the structure of the functional equation (\ref{eqfun}) the eigenvalues $T(u)$ must 
have zeroes which are organized in structures called $1-strings$ and $2-strings$.\\
1-strings are just single zeroes of real part $\lambda/2$ and imaginary part $0<v_k<-1/2\log q$ such that:
\be T(\lambda/2\pm iv_k)=0 \quad k=1,\ldots,m \ee
where $m$ denotes the number of 1-strings, while 2 strings are couples of zeroes sharing the same imaginary part, while their real 
part takes the values $\lambda/2\pm\lambda$, and we shall call their number $n$.\\ 
\begin{figure}
\protect\footnotesize
\setlength{\unitlength}{1pt}
\begin{picture}(140,190)(-110,10)
\multiput(80,20)(80,0){2}{\line(0,1){140}}
\put(60,90){\line(1,0){120}}
\put(30,145){\color{red}$m=2$}
\put(30,108){\color{green}$n=2$}
\multiput(80,60)(80,0){2}{\color{green}\circle*{6}}
\multiput(80,80)(80,0){2}{\color{green}\circle*{6}}
\multiput(80,100)(80,0){2}{\color{green}\circle*{6}}
\multiput(80,120)(80,0){2}{\color{green}\circle*{6}}
\put(60,80){$-\frac{\lambda}2$}
\put(121,80){$\frac{\lambda}2$}\put(120,87){\line(0,1){6}}
\put(162,80){$\frac32 \lambda$}
\put(120,40){\color{red}\circle*{6}}
\put(120,20){\color{red}\circle*{6}}
\put(120,140){\color{red}\circle*{6}}\put(190,135){$I_1=0$}
\put(120,160){\color{red}\circle*{6}}\put(190,155){$I_2=0$}
\end{picture}\caption{Example of the structure of zeroes labeled by the topological number \{0,0\}\label{zeri}}
\end{figure}
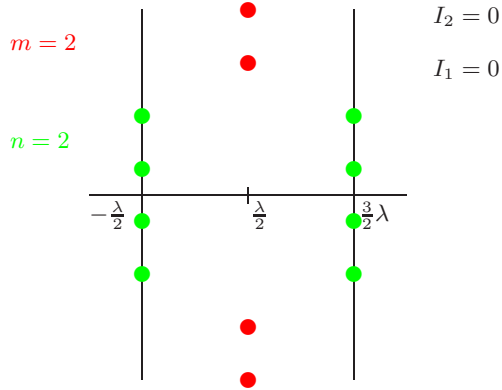

It is clear that a 1-string is a zero both for $T(u)$ and for $\mathcal{F}(u)$, therefore is will be convenient in our case to 
look for such zeroes in the expression for $\mathcal{F}$. Such zeroes coul in principle be found in either one of the three 
factors of which $\mathcal{F}$ is composed, but since (as we shall shortly see) $F_N$ and $S$ are going to be eliminated in the 
scaling limit 1-strings will essentially be zeroes of the $(1+d(u))$ term.
It is finally worth mentioning that for finite $N$ it is possible to give a characterization of the states (i.e. transfer matrix 
eigenvalues) in terms of a non increasing sequence of numbers $\{I_1,I_2,\ldots,I_m\}$ called \emph{quantum numbers} which express 
the position of 1-strings related to the position of 2-strings.\\
Each $I_k$ tells us how many 2-strings the $k$-th 1-string has to go through in order to reach its position in the pattern of 
zeroes starting from the configuration where all the 2-strings are heaped on the bottom, so that ordering  the imaginary part of 
the 1-strings $v_k$ into an  increasing sequence $\{v_k\}_{k=1}^m$ we have that the quantum numbers $\{I_k\}$ must necessarily 
arrange into a non increasing succession.\\
Clearly, the $I_k$ have to satisfy the following constraint:
\be I_k\leq n \ \forall k \ee
Such a characterization of the eigenvalues in terms of 1-strings and 2-strings also happens to give us a natural criterion for 
ordering the states, first of all we order the states by their increasing $m$ value, the ordering between equal $m$ states is done 
so that the state with all the 2-strings at the bottom of the tower comes first, and then each time a 2-string is pushed over a 
1-string the ``energy'' increases by one ``unit''.\\
We shall see that in the continuum limit a more natural set of quantum numbers will arise to describe the pattern of zeroes.  
\section{Excited TBA Equations}
In this section we are going to derive the excited state TBA equations for the $A_3$ model by solving (\ref{eqfun}). Considerable work has been done in the past on the excited states TBA, here we will essentially follow the work of \cite{pn}.\\
First of all let us recall the form of the Functional equation (\ref{eqfun}), we then define an $x$ coordinate in the following 
way:
\be \label{ics}
u=\frac{\lambda}2+i\frac{x}{4}, \qquad T_1(x) \defin T(u) 
\ee
where we are going to solve (\ref{eqfun}) for the following values of $x$:
\be \label{inter} 
x\in (2\log q,-2\log q) 
\ee 
for convenience we will rewrite (\ref{eqfun}) after applying a traslation:
\be u\rightarrow u-\frac{\lambda}2: \qquad 
T(u-\frac{\lambda}2 )~T(u+\frac{\lambda}2) = \mathcal{F}(u-\frac{\lambda}2) \ee
we then use (\ref{ics}) to write (\ref{eqfun}) in the following form:
\be u=\frac{\lambda}2+i\frac{x}{4} , \quad T_1(x+i\frac{\pi}{2})~T_1(x-i\frac{\pi}{2})
= \mathcal{F}(u-\frac{\lambda}2)=\mathcal{F}(i\frac{x}{4})
\defin \mathcal{F}_1(x) \label{funz}   \ee
At this point we could be tempted to follow the solution method used in \cite{pn} and try to Fourier-expand the logaritmic 
derivative of (\ref{funz}), anyway before being allowed to do so, we have to remove the zeroes of $T_1(x)$ in order to deal with 
an analytic function for which a Fourier expansion does make sense.\\
Now, if we consider what has been said in the previous section about the position of the zeroes, one observes that for real $x$, 
$|x|<-2\log q$, the function $T_1(x)$ which are due to the presence of 1-strings.\\
In order to reach our result we define the function
\be \label{pi}
p(x,v_k)=i\frac{\theta_1(\frac{i}{2} (x-4v_k),q^{2})}
{\theta_2(\frac{i}{2} (x-4v_k),q^{2})}
\ee

We observe that the $p$ function satisfies an equation which is similar to  $T_1(x)$:
\be \label{vincolo}
p(x+i\frac{\pi}{2},v_k) ~ p(x-i\frac{\pi}{2},v_k) =1
\ee
furthermore we observe that $p$ can be used to collect all the zeroes (for real $x$) of $T_1$ through the product:
\be \prod_{k=1}^m p(x,v_k)p(x,-v_k) \ee
so that we can assert that the function
\be
T_{\text{ANZ}}(x)\defin \displaystyle \frac{T_1(x)}
{\displaystyle\prod_{k=1}^m p(x,v_k)p(x,-v_k)}
=T_1(x) \prod_{k=1}^m
\frac{\theta_2(\frac{i}{2} (x-4v_k),q^{2})}{\theta_1(\frac{i}{2} (x-4v_k),q^{2})}~
\frac{\theta_2(\frac{i}{2}(x+4v_k),q^{2})}{\theta1(\frac{i}{2}(x+4v_k),q^{2})}
\ee
does not have zeroes for real $x$ $|x|< -2\log q$ (ANZ stands for analytic and not zero).\\
We then observe that as a consequence of (\ref{vincolo})$T_{ANZ}$ still satisfies the Functional equation:
\be
T_{\text{ANZ}}(x+i\frac{\pi}{2})~T_{\text{ANZ}}(x-i\frac{\pi}{2}) = 
\mathcal{F}_1(x)
\ee
so that now one is authorized to fourier-expand the logarithmic derivative of the above equation.Bymeans of some algebra one can 
determine $T_{ANZ}(x)$, and thus $T_1(x)$ to be:
\be
\log T_{1}(x)= \sum_{k=1}^m \log [p(x,v_k)p(x,-v_k)]+k *\log \mathcal{F}_1 + D
\ee 
Where $k(x)$ is a convolution kernel defined as:
\be k(x-y)= -\frac{1}{4\log q}\sum_{k=-\infty}^{\infty}\frac{e^{\frac{ik\pi (x-y)}{2\log q}}}{e^{-\frac{k\pi^2}{4\log 
q}}+e^{\frac{k\pi^2}{4\log q}}} \ee
The convolution kernel $k(x)$ can be computed  in terms of  Elliptic $\theta$ functions.
It has been computed  in \cite{pn} to have the following form:
\be k(x,q)=\frac{\theta_2(0,q^4)\theta_3(0,q^4)\theta_3(ix,q^4)}{2\pi\theta_2(ix,q^4)} \ee
Finally if we recall $\mathcal{F}(u)$'s definition we can write:
\be \log\mathcal{F}_1(x)= \log \mathcal{F}(i\frac{x}{4}) =
\log \Big( 1 +d(i\frac{x}{4}) \Big) 
+\log F_N(i\frac{x}{4})+\log S(i\frac{x}{4},\xi_L,\xi_R)\ee
so that remembering the expression for $F_N(i\frac{x}{4})$ we observe that its logarithm gives a contribution proportional to $N$ 
and thus can be identified with a \emph{Bulk Energy} term (which diverges in the thermodynamic limit $N\to\infty$), while 
$S(i\frac{x}{4},\xi_L,\xi_R)$ gives a contribution independent of $N$ which anyway diverges as we approach the critical regime 
($q\to 0$) and can thus be identified with a \emph{Surface Energy}.\\
Now, since in the next section we are going to deal with the continuum limit of $\log T(u)$ (which consists both of $N\to\infty$ 
and $q\to 0$), it is natural to conclude by defining a subtracted Energy so as to give rise to meaningful quantities in the 
continuum limit:
\be \label{rinorm}
\log T_{\text{finite}}(x) \defin \log T_1(x) 
- k *\log \big[ F_N(i\frac{x}{4})
S(i\frac{x}{4},\xi_L,\xi_R) \big]
\ee

and, more explicitly:
\be \label{dfinito}
\log T_{\text{finite}}(x)= 
\sum_{k=1}^m \log [p(x,v_k)p(x,-v_k)]+
k *\log \big(1+d(i\frac{x}{4})\big)
\ee

Where we observe that the constant of integration $D$ has finally disappeared in the subtraction, corresponding to a shift in the 
vacuum energy.


\section{Scaling Limit}
Taking a scaling limit of a lattice model essentially means considering its critical behaviour in the thermodynamical limit.\\
such a double limit ($N\to\infty, \ q\to 0$) can in principle be computed along infinite paths, however it can be shown \cite{pn} that there exists a prescription which allows us to obtain a meaningful limit for $\log D_{finite}$.\\ 
such a prescription has the form:
\be
q=t^{\frac{1}{2}}
\ee
\be
u=\frac{i}{4}(x+\log N)
\ee
where the size $N$ and the \emph{reduced temperature} $t$ satisfy:
\be N\to\infty\ee
\be N t^\nu=\mu \ee
$\nu$ is understood as the critical exponent of the correlation length, which for the $\mathbf{A}_n$ models in regime III of 
\cite{pn} is known to be:
\be \nu=\frac{n+1}{4} \ee
The $\mu$ parameter plays the role a regulator for the continuum system, and it will be used to generate a massive RG flow 
connecting the UV ($\mu=0$) and IR ($\mu=+\infty$) fixed points.\\
Such a regulator can be thought of as arising from the product of a mass $m$ and a length $R$:
\be\mu=\frac{1}{4}mR \ee
being
\be R=\lim_{N\to\infty,l\to 0}N l \ee
\be m=\lim_{t\to 0, l\to 0} \frac{t^\nu}{l} \ee
and $l$ is understood as the lattice spacing.\\  
Our first goal is to build a continuum version of $\d(u,q)$ which we will use to expand the continuum transfer matrix as a series 
whose coefficients will be the integrals of motion, we will then use the information we will gain to dicuss the full continuum 
expression for $\log T_{\text{finite}}(x)$.\\
The limit we are essentially interested in computing is the following:
\be \hat d(x):=\lim_{N\to\infty}d\Big(\frac{i}{4}(x+\log N),\big(\frac{\mu}{N}\big)^{\frac{1}{2}}\Big)\ee
From now on we shall fix the right boundary to $a_R=1$ so that already before going into the scaling limit the R boundary term is fixed to:
\be \mathcal{B}_R=-1 \ee
Therefore from now on the only surviving boundary height $a_L$ will be simply called $a$ and we shall realize in what will follow that $a=s$ where $s$
is the Kac label of $\Delta_{r,s}$, and of course in this case $r=a_R=1$.\\
In order to achieve the correct scaling behaviour for the boundary term one has to postulate 
the following scaling behaviour for the boundary parameter $\xi$ (we are dropping the subscript L):
\be \xi\sim\xi^* +\frac{1}{4}\log N \ee
It is also useful to split the boundary term as:
\be \mathcal{B}=e^{i\pi a}\mathcal{B}^+\mathcal{B}^- \ee
so that $\mathcal{B}^+$ depends only on $u+\xi$ and similarly $\mathcal{B}^-$ depends only on $u-\xi$. 
Proceeding further one finds:
\be \hat{\mathcal{ B}}^+=\frac{1-\mu^{\frac{1}{2}}e^{\frac{x}{2}}e^{-2i(\xi^*+\frac{\pi}{4})}}{1-\mu^{\frac{1}{2}}e^{\frac{x}{2}}e^{-2i(\xi^*+a\frac
{\pi}{4})}}\frac{1-\mu^{\frac{1}{2}}e^{\frac{x}{2}}e^{-2i(\xi^*-\frac{\pi}{4})}}{1-\mu^{\frac{1}{2}}e^{\frac{x}{2}}e^{-2i(\xi^*-(a
+2)\frac{\pi}{4})}}\ee
\be \hat{\mathcal{B}}^-=\frac{1-\mu^{\frac{1}{2}} 
e^{\frac{x}{2}}e^{-2i(\frac{\pi}{4}-\xi^*)}}{1-\mu^{\frac{1}{2}} 
e^{\frac{x}{2}}e^{2i(a\frac{\pi}{4}+\xi^*)}}\frac{1-\mu^{\frac{1}{2}} 
e^{\frac{x}{2}}e^{2i(\frac{\pi}{4}+\xi^*)}}{1-\mu^{\frac{1}{2}} e^{\frac{x}{2}}e^{2i((a+2)\frac{\pi}{4}+\xi^*)}}\ee
One then notices that in the UV limit $\mu\to 0 $\ the boundary $\hat{\mathcal{B}}$ 
flows into
\be \hat{\mathcal{B}}\sim e^{i\pi a}\ee 
and takes the value $-1$ for $a=1,3$ and $+1$ for $a=2$.\\
After carrying out the calculation one finds the following continuum version of $\d$ :
\be \hat d(x,\mu)= e^{-8\mu\cosh(x+\log\mu)}\tanh^2\Big(\frac{x+\log\mu}{2}\Big)\big(\hat{\mathcal{B}_L
}\hat{\mathcal{B}_R}\big) \ee
One should at this point discuss the scaling limit of the convolution term in $\log T_{\text{finite}}(x)$, such a calculation 
turns out to yield:
\be\label{cont} \lim_{N\to\infty, q\to 0}k *\log \Big(1+d\Big(\frac{i}{4} (x+\log N)\Big)\Big)=\int_{-\infty}^{+\infty}\frac{1}{2\pi\cosh 
(x-y)}\log (1+\hat{d}(y))dy \ee
We can now deal with the scaling limit of the excitations.\\
As we approach the scaling limit the 1-strings will have the following asymptotic behaviour ($y_k$ is the finite part):
\be 4v_k\sim y_k+\log N \ee
The 1-string term will then become:
\be \sum_{k=1}^m\log(p(x,v_k)p(x,-v_k))\sim\sum_{k=1}^m\log\Big(\tanh\big(\frac{x-y_k+\log\mu}{2}\big)\tanh
\big(\frac{x+y_k+\log\mu}{2}\big)\Big) \ee

\section{Expansion}
\subsection{Study of the Ground State}
We first of all begin by studying the behaviour of the continuum ground state eigenvalue of the transfer matrix; in the previous 
section we have shown it takes the following form:
\be \label{giterm}\log\hat D(x)=\int_{-\infty}^{+\infty}\frac{dy}{2\pi}\frac{\log(1+\hat d(y))}{\cosh(x-y)} \ee
such an expression, following the spirit of \cite{tba} has to be expanded in the following series:
\be \log\hat D(x)=-\sum_{n=1}^{\infty}C_nI_{2n-1}(\mu)e^{(2n-1)x} \ee
which in our case yields the following expression for the vacuum integrals of motion:
\be \label{groundint} C_nI^{vac}_{2n-1}(\mu)=\frac{(-1)^n}{\pi}\int_{-\infty}^{+\infty}dy e^{-(2n-1)y}\log(1+\hat d(y)) \ee 
we're now going to study this expression in the UV and IR limits.\\
We start by observing that the above expression can be manipulated into the form:
\be\label{cibess} C_nI_{2n-1}^{vac}=\frac{(-1)^n \mu^{2(2n-1)}}{\pi 4^{(2n-1)}}\sum_{k=1}^\infty 
\frac{(-1)^{k+1}}{k^{2n}}\int_0^\infty \frac{dt}{t} 
t^{(2n-1)}e^{-(t+\frac{16k^2\mu^2}{t})}\Big(\frac{t-4k\mu}{t+4k\mu}\Big)^{2k}\Big(\hat{\mathcal{B}_R}\hat{\mathcal{B}_L}\Big
)^k \ee
Where we once we fix the right boundary to $a_R=1$ $\hat{\mathcal{B}_R}$ disappears from the equations.\\
One at this point decides to get rid of $\xi^*$ fixing it to zero. Actually one could keep it, and let it scale once again as 
$\mu\to 0$ so that, for a suitable $\Lambda$:
\be \mu^\Lambda e^{4i\xi^*}=\tau \ee
The parameter $\tau$ would then generate a flow between different conformal boundary conditions, see for example \cite{changrim}.\\ 
Now, by means of standard techniques one can prove that the following inequality holds for all values of $a$:
\be\begin{split} & | C_nI^{vac}_{2n-1}(\mu)|\leq \frac{2\mu^{(2n-1)}}{\pi}\sum_{k=1}^\infty 
\frac{1}{k}\textrm{K}_{1-2n}(8k\mu)\end{split} \ee
where the $K_l(z)$ are the modified Bessel functions of the second kind. A study of the large $\mu$ asymptotics of the above 
series allows one to conclude that the ground states of the integrals of motion decay exponentially in the IR limit.\\
We are now ready to move our attention to the UV asymptotic behaviour.\\
In the limit $\mu\to 0$ it is not difficult to show that: 
\be C_nI_{2n-1}^{vac}(\mu)\sim \frac{(-1)^{n+1}}{\pi 4^{(2n-1)}}\Gamma(2n-1)\textrm{Li}_{2n}(e^{i\pi a}) \ee
where 
\be \mathrm{Li}_{\nu}(z)=\sum_{k=1}^{\infty}\frac{z^k}{k^\nu} \ee
is the Polylogarithm function. It is worth spending a word to observe that considering the energy $I_1$ one finds a dilogarithm of 
a phase, actually it is well known that the central charge is usually proportional to a dilogarithm. Furthermore one notices that 
all the vacuum expectation values of the integrals of motion are proportional to polylogarithms, this seems to be a rather general 
structure.One could ask himself if such polylogarithms satisfy sum rules similar to those holding for dilogarithms.\\

\subsection{Excited States}
We now ask ourselves what is the behaviour of the excitation terms in the UV and IR and compare it with the 
ground states.\\
 Now we want to expand the 1-string term.\\
The expansion is readily obtained if one considers the following identity:
\be\label{esp1}\begin{split} &\log\Bigg(\frac{1-t}{1+t}\Bigg)=-2\sum_{k=1}^\infty\frac{t^{2k-1}}{2k-1}      \end{split}\ee
and for example writes:
\be\begin{split} &\log\tanh\frac{x-y_k+\log\mu}{2}=i\pi-2\sum_{n=1}^\infty e^{(2n-1)x}\frac{e^{-(2n-1)(y_k-\log\mu)}}{2n-1}\end{split}\ee
Clearly one obtains the following result:
\be 
\Big(1-string\Big)=-\sum_{n=1}^\infty\frac{2e^{(2n-1)x}}{2n-1}\Big(\sum_{k=1}^m\big(e^{-(2n-1)y_k}+\mu^{2(2n-1)}e^{(2n-1)y_k}\big)
\Big) \ee
In order to proceed further it is necessary to have a closer look at the asymptotic behaviour of the $y_k$.\\
Let us recall the equation satisfied by the $y_k$:
\be \hat d\big(y_k-i\frac{\pi}{2}\big)=-1 \ee
so that by taking the logarithm of both sides, using the expression for $\hat d$ given previously we have:
\be -4\mu(e^{-2i\lambda} e^{ 
y_k}\mu-e^{2i\lambda}e^{-y_k}\mu^{-1})+2\log\Big(\frac{1-e^{2i\lambda}\mu^{-1}e^{-y_k}}{1+e^{2i\lambda}\mu^{-1}e^{-y_k}}\Big)+
\log(\hat{\mathcal{B}}_L\hat{\mathcal{B}}_R)=i\pi n_k \ee
where the $n_k$ are odd numbers.\\
We now introduce the following function:
\be g_k(\mu):=\mu e^{y_k} \ee
we are now interested in expressing the $y_k$ equation in terms of this new function $g_k$, in order to do so we first rewrite the 
boundary term as:
\be \hat{\mathcal{B}}(y_k-i\frac{\pi}{2})=e^{i\pi a}\frac{(g_k+i)^2}{(g_k+i)^2-2ig_k(1+\cos(\pi a ))}\ee 
so that the $y_k$ equation becomes simpler and reads:
\be 4i\mu \frac{(g_k^2-1)}{g_k}+\log\Big(\frac{(g_k-i)^2}{(g_k+i)^2-2ig_k(1+\cos(\pi a))}\Big)=i\pi(n_k+1-a)\ee
this equation gives us the inverse function $\mu(g_k)$, and we are interested in its behaviour as $\mu\to 0$and $\mu\to+\infty$. 
In order to reach our goal it suffices to pick the branch of the $\mu(g_k)$ function which passes through the origin, $g_k\to 0$ 
corresponds to the UV limit, whereas $g_k\to 1$ is the IR limit.\\
Expanding $\mu$ around $g_k=0$ we obtain:
\be \mu\sim\frac{\pi}{4}(a-1-n_k)g_k+O(g_k^2) \ee 
If we now decide to rewrite the 1-string term as follows:
\be 
(1-string)=-\sum_{n=1}^\infty\frac{2e^{(2n-1)x}}{2n-1}\mu^{\eta(2n-1)}\sum_{k=1}^m\Big(g_k^{(2n-1)}+\frac{1}{g_k^{(2n-1)}}\Big) 
\ee
we conclude by applying the above UV expansion that the expression
\be \mu^{(2n-1)}\Big(g_k^{(2n-1)}+\frac{1}{g_k^{(2n-1)}}\Big) \ee
is UV limited, so that in this limit the excitations have the same scaling behaviour as the ground state.\\
On the other hand the IR excitations cannot avoid to grow faster than the ground state term, this observation united to the fact 
that $g_k\to 1$ dictates the particular structure of the IR spectrum.\\
\subsection{Full Expansion}
We finally have arrived at the point of writing the analytic expression for all the integrals of motion of the model, such 
expression reads:
\be\begin{split} C_nI_{2n-1}(\mu)=&\frac{2\mu^{(2n-1)}}{2n-1}\sum_{k=1}^m\Big(g_k^{(2n-1)}+\frac{1}{g_k^{(2n-1)}}\Big)+\\ 
&+\frac{(-1)^n}{\pi}\int_{0}^{+\infty}dye^{-(2n-1)y}\log(1+\hat d(y,\mu)) \end{split}\ee 
In the limit $\mu\to\infty, \ g_k\to 1$ the ground state drops off exponentially, so that considering only the excitations it is 
immediate to realize that:
\be  C_nI_{2n-1}(\mu)\sim \frac{4m\mu^{(2n-1)}}{2n-1} \ee
this spectrum happens to be integrally spaced, and looses all memory of the quantum numbers aside from the length of the sequence 
of the $n_k$.\\
It is worth remarking that this result should not surprise us very much since it is very similar to what has been achieved in 
\cite{torelloni} for the $\mathbf{A}_4$ model corresponding to the tricritical Ising Model  universality class.\\
In the UV limit it is clear from what we said so far that the integrals of motion have the following behaviour:
\be\begin{split} C_nI_{2n-1}(\mu)\sim& \Big(\frac{\pi}{4}\Big)^{(2n-1)}\Big(\frac{2}{2n-1}\sum_{k=1}^m 
\Big(a-1-n_k\Big)^{(2n-1)}+\\&+(-1)^{n}e^{i\pi a}((1-e^{i\pi a})2^{-2n}-1)\Gamma(2n-1) \frac{\zeta(2n)}{\pi^{2n}} 
\Big)\end{split} \ee
For $a=1$ the energy takes the following form
\be C_1I_1(\mu)\sim \pi\Big(\frac{1}{2}\sum_{k=1}^m (-n_k) -\frac{1}{48}\Big) \ee
For $a=2$ we have:
\be C_1I_1(\mu)\sim \pi\Big(\frac{1}{2}\sum_{k=1}^m (1-n_k) +\frac{1}{24}\Big) \ee
For $a=3$ we have:
\be C_1I_1(\mu)\sim \pi\Big(\frac{1}{2}\sum_{k=1}^m (2-n_k) -\frac{1}{48}\Big) \ee
It is readily recognized that these formulae are in agreement respectively with the $h=0$,$h=1/16$ and $h=1/2$ sectors of the 
minimal model $\mathcal{M}_{3,4}$ if we choose:
\be C_1=\pi \ee
One notices that  in the vacuum sector $m$ must be even whereas in the $1/2$ sector $m$ must be odd for trivial reasons. In the $1/16$ sector $m$ must be odd but this fact is less trivial to understand from the formula, let us simply say that $m-$parity is fixed in the sector and is odd because for the highest weight state the only possible quantum number must satisfy $1-n_1=0$.\\
 We will understand better the structure of the quantum numbers in the next section.\\
If we now want to compute the constants $C_2,C_3$ in the vacuum sector of the model this can be done by using the explicit 
expressions for the integrals of motion, which can be found in \cite{blz}:
\be \mathbf{I}_{1}=L_0-\frac{c}{24} \ee
\be \mathbf{I}_{3}= 2\sum_{n=1}^{\infty}L_{-n}L_n+L_0^2-\frac{c+2}{12}L_0+\frac{c(5c+22)}{2880} \ee
\be \begin{split}\mathbf{I}_{5}&=\sum_{m,n,p\in\mathbb{Z}}\delta_{m+n+p,\ 0}:L_m L_n L_p:+\frac{3}{2}\sum_{n=1}^\infty L_{1-2n}L_{2n-1}+\\
&+\sum_{n=1}^\infty\Bigg(\frac{11+c}{6}n^2-\frac{c}{4}-1\Bigg)L_n L_{-n}-\frac{c+4}{8}L_0^2+\frac{(c+2)(3c+20)}{576}L_0+\\
&-\frac{c(3c+14)(7c+68)}{290304} \end{split} \ee
where the $:\ :$ denotes \emph{Conformal Normal Ordering} which can be obtained by arranging all the $L_n$ in an increasing sequence with respect to 
$n$.\\
So that we have the following vacuum expectation values:
\be \big<0\big|\mathbf{I}_3\big|0\big>=\frac{49}{11520} \ee
\be \big<0\big|\mathbf{I}_5\big|0\big>=-\frac{4433}{2322432} \ee
which gives us the following equations:
\be C_2\big<0\big|\mathbf{I}_3\big|0\big>=\frac{\pi^3\Gamma(3)(1-2^{-3})}{4^3 90} \ee
\be C_3\big<0\big|\mathbf{I}_3\big|0\big>=-\frac{\pi^5\Gamma(5)(1-2^{-5})}{4^5 945} \ee
so that we readily get:
\be C_2=\frac{\pi^3}{14}\ee
\be C_3=\frac{9}{715}\pi^5 \ee
This values for the vacuum constants $C_n$ actually can be extracted from \cite{blz1}, so that in 
general we have for the vacuum sector the following expression:
\be C_n=\frac{3^n 4^{2-3n}\pi^{-\frac{1}{2}+2n}\Gamma(4n-2)}{n!\Gamma(3n-\frac{1}{2})}  \ee
Actually these values of the $C_n$ are computed from the vacuum sector, but direct calculation allows one to verify that they are 
independent of the sector.\\
We stress that the above formulas describe exactly the conformal spectrum of the minimal model $\mathcal{M}_{3,4}$.\\

 \begin{table}[p]
\protect\footnotesize
\centering
\begin{tabular}{|ccc|}
\hline CFT State &Fermionic State  & TBA State \\
\hline
\hline $\big| 0\big>$ &$\big|0\big>$  & $()$\\
\hline $2L_{-2}\big|0\big>$ & $\psi_{-\frac{3}{2}}\psi_{-\frac{1}{2}}\big|0\big>$ & $(-1,-3)$\\
\hline $L_{-3}\big|0\big>$ & $\psi_{-\frac{5}{2}}\psi_{-\frac{1}{2}}\big|0\big>$ & $(-1,-5)$\\
\hline $\frac{5}{7}L_{-4}\big|0\big>-\frac{6}{7}L_{-2}^2\big|0\big>$ & $\psi_{-\frac{5}{2}}\psi_{-\frac{3}{2}}\big|0\big>$ & $(-3,-5)$ \\
 $\frac{3}{7}L_{-4}\big|0\big>+\frac{2}{7}L_{-2}^2\big|0\big>$ & $\psi_{-\frac{7}{2}}\psi_{-\frac{1}{2}}\big|0\big>$ & $(-1,-7)$\\
 \hline $\frac{3}{7}L_{-5}\big|0\big>-\frac{4}{7}L_{-3}L_{-2}\big|0\big>$ & $\psi_{-\frac{7}{2}}\psi_{-\frac{3}{2}}\big|0\big>$ & $(-3,-7)$\\
 $\frac{2}{7}L_{-5}\big|0\big>+\frac{2}{7}L_{-3}L_{-2}\big|0\big>$ & $\psi_{-\frac{9}{2}}\psi_{-\frac{1}{2}}\big|0\big>$ & $(-1,-9)$\\
 \hline $\frac{5}{14}L_{-6}\big|0\big>+\frac{3}{7}L_{-4}L_{-2}\big|0\big>-\frac{23}{56}L_{-3}^2\big|0\big>$ & $\psi_{-\frac{7}{2}}\psi_{-\frac{5}{2}}\big|0\big>$ & $(-5,-7)$ \\
 $\frac{1}{4}L_{-6}\big|0\big>-\frac{1}{2}L_{-4}L_{-2}\big|0\big>+\frac{1}{16}L_{-3}^2\big|0\big> $ & $\psi_{-\frac{9}{2}}\psi_{-\frac{3}{2}}\big|0\big>$ & $(-3,-9)$\\
 $\frac{5}{28}L_{-6}\big|0\big>+\frac{3}{14}L_{-4}L_{-2}\big|0\big>+\frac{5}{112}L_{-3}^2\big|0\big>$ & $\psi_{-\frac{11}{2}}\psi_{-\frac{1}{2}}\big|0\big>$ & $(-1,-11)$\\
 \hline
 \end{tabular}
  \caption{U.V. state correspondence CFT$\longrightarrow$TBA for the  $h=0$ sector}
\label{CFTTBA3}
\end{table}
 \begin{table}[p]
\protect\footnotesize
\centering
\begin{tabular}{|ccc|}
 \hline CFT State & Fermionic State  & TBA State \\
 \hline
 \hline $\big|1/2\big>$ & $\big|0\big>$ & $(1)$ \\
 \hline $L_{-1}\big|1/2\big>$ & $\psi_{-\frac{3}{2}}\big|0\big>$ & $(-1)$ \\
 \hline $\frac{2}{3}L_{-2}\big|1/2\big>$ & $\psi_{-\frac{5}{2}}\big|0\big>$ & $(-3)$ \\
 \hline $\frac{1}{2}L_{-3}\big|1/2\big>$ & $\psi_{-\frac{7}{2}}\big|0\big>$ & $(-5)$ \\
 \hline $\frac{1}{4}L_{-4}\big|1/2\big>+\frac{1}{8}L_{-3}L_{-1}\big|1/2\big>$ & $\psi_{-\frac{9}{2}}\big|0\big>$ & $(-7)$\\
 $\frac{3}{4}L_{-4}\big|1/2\big>-\frac{5}{8}L_{-3}L_{-1}\big|1/2\big>$ & $\psi_{-\frac{5}{2}}\psi_{-\frac{3}{2}}\psi_{-\frac{1}{2}}\big|0\big>$ & $(1,-1,-3)$\\
 \hline $\frac{3}{16}L_{-5}\big|1/2\big>+\frac{1}{8}L_{-4}L_{-1}\big|1/2\big>$ & $\psi_{-\frac{11}{2}}\big|0\big>$ & $(-9)$ \\
 $\frac{7}{16}L_{-5}\big|1/2\big>-\frac{3}{8}L_{-4}L_{-1}\big|1/2\big>$ & $\psi_{-\frac{7}{2}}\psi_{-\frac{3}{2}}\psi_{-\frac{1}{2}}\big|0\big>$ & $(1,-1,-5)$\\
 \hline $\frac{1}{8}L_{-6}\big|1/2\big>+\frac{3}{32}L_{-5}L_{-1}\big|1/2\big>+\frac{1}{36}L_{-4}L_{-2}\big|1/2\big>$ & $\psi_{-\frac{13}{2}}\big|0\big>$ & $(-11)$\\
 $\frac{1}{4}L_{-6}\big|1/2\big>-\frac{5}{16}L_{-5}L_{-1}\big|1/2\big>+\frac{1}{18}L_{-4}L_{-2}\big|1/2\big>$ & $\psi_{-\frac{9}{2}}\psi_{-\frac{3}{2}}\psi_{-\frac{1}{2}}\big|0\big>$ & $(1,-1,-7)$\\
 $\frac{3}{8}L_{-6}\big|1/2\big>+\frac{9}{32}L_{-5}L_{-1}\big|1/2\big>-\frac{13}{36}L_{-4}L_{-2}\big|1/2\big>$ & $\psi_{-\frac{7}{2}}\psi_{-\frac{5}{2}}\psi_{-\frac{1}{2}}\big|0\big>$ & $(1,-3,-5)$\\
 \hline                 
 \end{tabular}
 \caption{U.V. state correspondence CFT$\longrightarrow$TBA for the $h=1/2$ sector }
\label{CFTTBA4}
\end{table}

 \begin{table}[p]
 \protect\footnotesize
\centering
 \begin{tabular}{|ccc|} 
 \hline  CFT State  &Fermionic State  & TBA State \\
 \hline
 \hline $\big|1/16\big>_{-}$ & $\sqrt{2}\psi_0\big|1/16\big>_{+}$ & (1)\\
 \hline $2\sqrt{2}L_{-1}\big|1/16\big>_{-}$ & $\psi_{-1}\big|1/16\big>_{+}$ & (-1)\\
 \hline $\sqrt{2}L_{-2}\big|1/16\big>_{-}$ & $\psi_{-2}\big|1/16\big>_{+}$ & (-3)\\
 \hline $\frac{8}{7}L_{-3}\big|1/16\big>_{-}-\frac{12}{7}L_{-2}L_{-1}\big|1/16\big>_{-}$ & $\sqrt{2}\psi_{-2}\psi_{-1}\psi_{0}\big|1/16\big>_{+}$ & (1,-1,-3)\\
 $\frac{2\sqrt{2}}{7}L_{-3}\big|1/16\big>_{-}+\frac{4\sqrt{2}}{7}L_{-2}L_{-1}\big|1/16\big>_{-}$ & $\psi_{-3}\big|1/16\big>_{+}$ & (-5)\\
 \hline $\frac{5}{8}L_{-4}\big|1/16\big>_{-}-L_{-3}L_{-1}\big|1/16\big>_{-}$ & $\sqrt{2}\psi_{-3}\psi_{-1}\psi_{0}\big|1/16\big>_{+}$ & (1,-1,-5)\\
 $\frac{3}{8\sqrt{2}}L_{-4}\big|1/16\big>_{-}+\frac{1}{\sqrt{2}}L_{-3}L_{-1}\big|1/16\big>_{-}$ & $\psi_{-4}\big|1/16\big>_{+}$ & (-7)\\
 \hline $\frac{1}{2}L_{-5}\big|1/16\big>_{-}+\frac{3}{2}L_{-4}L_{-1}\big|1/16\big>_{-}-L_{-3}L_{-2}\big|1/16\big>_{-}$ & $\sqrt{2}\psi_{-3}\psi_{-2}\psi_{0}\big|1/16\big>_{+}$ & (1,-3,-5)\\
 $\frac{9}{28}L_{-5}\big|1/16\big>_{-}-\frac{29}{28}L_{-4}L_{-1}\big|1/16\big>_{-}+\frac{3}{14}L_{-3}L_{-2}\big|1/16\big>_{-}$ & $\sqrt{2}\psi_{-4}\psi_{-1}\psi_{0}\big|1/16\big>_{+}$ & (1,-1,-7)\\
 $\frac{3}{14\sqrt{2}}L_{-5}\big|1/16\big>_{-}+\frac{9}{14\sqrt{2}}L_{-4}L_{-1}\big|1/16\big>_{-}+\frac{1}{7\sqrt{2}}L_{-3}L_{-2}\big|1/16\big>_{-}$ & $\psi_{-5}\big|1/16\big>_{+}$ & (-9)\\
 \hline$\frac{19}{64\sqrt{2}}L_{-6}\big|1/16\big>_{-}-\frac{5}{8\sqrt{2}}L_{-5}L_{-1}\big|1/16\big>_{-}+\frac{1}{\sqrt{2}}L_{-4}L_{-2}\big|1/16\big>_{-}-\frac{9}{16\sqrt{2}}L_{_3}^2\big|1/16\big>_{-}$ & $\psi_{-3}\psi_{-2}\psi_{-1}\big|1/16\big>_{+}$ & 
(-1,-3,-5)\\
 $\frac{409}{896}L_{-6}\big|1/16\big>_{-}+\frac{81}{112}L_{-5}L_{-1}\big|1/16\big>_{-}-\frac{3}{14}L_{-4}L_{-2}\big|1/16\big>_{-}-\frac{59}{224}L_{-3}^2\big|1/16\big>_{-}$ & 
$\sqrt{2}\psi_{-4}\psi_{-2}\psi_{0}\big|1/16\big>_{+}$ & (1,-3,-7)\\
 $\frac{5}{32}L_{-6}\big|1/16\big>_{-}-\frac{3}{4}L_{-5}L_{-1}\big|1/16\big>_{-}+\frac{1}{8}L_{-3}^2\big|1/16\big>_{-}$&$\sqrt{2}\psi_{-5}\psi_{-1}\psi_{0}\big|1/16\big>_{+}$&(1,-1,-9)\\
 $\frac{69}{448\sqrt{2}}L_{-6}\big|1/16\big>_{-}+\frac{29}{56\sqrt{2}}L_{-5}L_{-1}\big|1/16\big>_{-}+\frac{1}{7\sqrt{2}}L_{-4}L_{-2}\big|1/16\big>_{-}+\frac{1}{112\sqrt{2}}L_{-3}^2\big|1/16\big>_{-}$ & $\psi_{-6}\big|1/16\big>_{+}$ & 
(-11)\\
 \hline 
 \end{tabular}
 \caption{U.V. state correspondence CFT$\longrightarrow$TBA for the $h=1/16$ sector}
\label{h16c12}
\end{table}

\begin{table}[p]

\centering
 \begin{tabular}{|ccc|} 
 \hline  TBA State  &$I_3$ Eigenvalue  & $I_5$ Eigenvalue \\
\hline$()$& $\frac{49}{ 11520}$& $\frac{-4433}{ 2322432}$\\
 \hline$(-1, -3)$& $\frac{47089}{ 11520}$& $\frac{17581135}{ 2322432}$\\
     \hline$(-1, -5)$& $\frac{211729}{ 11520}$& $\frac{225292639}{2322432}$\\
     \hline$(-3, -5)$& $\frac{255409}{ 11520}$& $\frac{242734063}{2322432}$\\
    \hline  $(-1, -7)$& $\frac{577969}{ 11520}$& $\frac{1211381743}{2322432}$\\
  \hline   $(-1, -9)$& $\frac{1226449}{ 11520}$& $\frac{4255847167}{2322432}$\\ 
  \hline  $(-3, -7)$& $\frac{621649}{ 11520}$& $\frac{1228823167}{2322432}$\\
    \hline $(-5, -7)$& $\frac{786289}{ 11520}$& $\frac{1436534671}{2322432}$\\
  \hline   $(-3, -9)$& $\frac{1270129}{ 11520}$& $\frac{4273288591}{2322432}$\\ 
  \hline  $(-1, -11)$& $\frac{2237809}{ 11520}$& $\frac{11607335311}{2322432}$\\
 \hline 
 \end{tabular}
 \caption{Table of eigenvalues in the vacuum sector}
\label{eig0}
\end{table}

\begin{table}[p]

\centering
 \begin{tabular}{|ccc|} 
 \hline  TBA State  &$I_3$ Eigenvalue  & $I_5$ Eigenvalue \\
\hline $(1)$& $\frac{1729}{ 11520}$& $\frac{67639}{ 2322432}$\\
 \hline $(-1)$& $\frac{45409}{ 11520}$& $\frac{17509063}{2322432}$\\
  \hline $(-3)$& $\frac{210049}{ 11520}$& $\frac{225220567}{2322432}$\\
   \hline $(-5)$& $\frac{576289}{ 11520}$& $\frac{1211309671}{2322432}$\\
    \hline $(-7)$& $\frac{1224769}{ 11520}$& $\frac{4255775095}{2322432}$\\
     \hline $(1, -1, -3)$& $\frac{257089}{ 11520}$& $\frac{242806135}{2322432}$\\
      \hline $(-9)$& $\frac{2236129}{ 11520}$& $\frac{11607263239}{2322432}$\\
       \hline $(1, -1, -5)$& $\frac{623329}{ 11520}$& $\frac{1228895239}{2322432}$\\
        \hline $(-11)$& $\frac{3691009}{ 11520}$& $\frac{26759824663}{2322432}$\\
         \hline $(1, -1, -7)$& $\frac{1271809}{ 11520}$& $\frac{4273360663}{2322432}$\\
          \hline $(1, -3, -5)$& $\frac{787969}{ 11520}$& $\frac{1436606743}{ 2322432}$\\
\hline 
 \end{tabular}
 \caption{Table of eigenvalues in the 1/2 sector}
\label{eig05}
\end{table}

\begin{table}[p]

\centering
 \begin{tabular}{|ccc|} 
 \hline  TBA State  &$I_3$ Eigenvalue  & $I_5$ Eigenvalue \\
\hline$(1)$& $\frac{-7}{ 1440}$& $\frac{143}{ 72576}$\\
\hline $(-1)$& $\frac{1673}{ 1440}$& $\frac{72215}{ 72576}$\\
  \hline$(-3)$& $\frac{13433}{ 1440}$& $\frac{2306447}{ 72576}$\\
  \hline $(1, -1, -3)$& $\frac{15113}{1440}$& $\frac{2378519}{ 72576}$\\
   \hline $(-5)$&$\frac{45353}{ 1440}$& $\frac{17513639}{72576}$\\
   \hline  $(1, -1, -5)$& $\frac{47033}{ 1440}$& $\frac{17585711}{72576}$\\
    \hline  $(-7)$& $\frac{107513}{ 1440}$& $\frac{73801871}{72576}$\\
    \hline   $(1, -3, -5)$& $\frac{58793}{ 1440}$& $\frac{19819943}{72576}$\\
      \hline  $(1, -1, -7)$& $\frac{109193}{ 1440}$& $\frac{73873943}{72576}$\\
       \hline  $(-9)$& $\frac{209993}{1440}$& $\frac{225225143}{72576}$\\
     \hline     $(-1, -3, -5)$& $\frac{60473}{1440}$& $\frac{19892015}{72576}$\\
         \hline  $(1, -3, -7)$& $\frac{120953}{1440}$& $\frac{76108175}{72576}$\\
         \hline   $(1, -1, -9)$& $\frac{211673}{1440}$& $\frac{225297215}{72576}$\\
       \hline      $(-11)$& $\frac{362873}{ 1440}$& $\frac{560432015}{ 72576}$\\
\hline 
 \end{tabular}
 \caption{Table of eigenvalues in the 1/16 sector} 
\label{eig16}
\end{table}

\section{Fermionic modes and TBA quantum numbers }
Actually the quantum numbers $n_k$ themselves have a very simple interpretation in terms of fermionic modes. \\
To understand this one needs only to remember that the stress energy tensor for the Ising model is built out of the fermion field as:
\be T(z):=\frac{1}{2}:\psi(z)\partial\psi(z): \ee
So that by introducing the well known mode expansion
\be i\psi(z)=\sum_n \frac{\psi_n}{z^{n+\frac{1}{2}}}\ee
one gets
\be L_{n}=\frac{1}{2}\sum_{k}(k+\frac{1}{2}):\psi_{n-k}\psi_{k}: \ee
Where the fermionic modes will have a half integer index when we will be working in the $0,1/2$ sectors, whereas the index will be integer in the twisted $1/16$ sector.\\
It is then just a matter of unraveling the normal ordering and using the fermionic algebra $\{\psi_n,\psi_m\}=\delta_{n+m, 0}$ to work out the fermionic expression for the eigenvectors of the integrals of motion.\\
If we consider for example the sixth level of descendance in the vacuum sector we have:
\be 20L_{-6}\big|0\big>+24L_{-4}L_{-2}\big|0\big>+5L_{-3}^2\big|0\big>=112\psi_{-\frac{11}{2}}\psi_{-\frac{1}{2}}\big|0\big> \ee
\be 4L_{-6}\big|0\big>-8L_{-4}L_{-2}\big|0\big>+L_{-3}^2\big|0\big>=16\psi_{-\frac{9}{2}}\psi_{-\frac{3}{2}}\big|0\big> \ee
\be 20L_{-6}\big|0\big>+24L_{-4}L_{-2}\big|0\big>-23L_{-3}^2\big|0\big>=56\psi_{-\frac{7}{2}}\psi_{-\frac{5}{2}}\big|0\big> \ee
So that comparing with table \ref{CFTTBA3} we see that $\frac{n_k}{2}$ are simply the labels of the fermionic modes, and we can 
easily either guess or explicitly work out straightforwardly the form of all the other eigenstates which is obvious aside from a 
normalization.\\
Similarly in the $1/2$ sector one has at level 6:
\be 36L_{-6}\big|1/2\big>+27L_{-5}L_{-1}\big|1/2\big>+8L_{-4}L_{-2}\big|1/2\big>=288\psi_{-\frac{13}{2}}\big|0\big> \ee
\be 
36L_{-6}\big|1/2\big>-45L_{-5}L_{-1}\big|1/2\big>+8L_{-4}L_{-2}\big|1/2\big>=144\psi_{-\frac{9}{2}}\psi_{-\frac{3}{2}}\psi_{-\frac
{1}{2}}\big|0\big> \ee
\be 108L_{-6}\big|1/2\big>+81L_{-5}L_{-1}\big|1/2\big>-26L_{-4}L_{-2}\big|1/2\big>=72\psi_{-\frac{7}{2}}\psi_{-\frac{5}{2}}\psi_{- 
\frac{1}{2}}\big|0\big> \ee
so that the fermionic modes have indexes which are simply $\frac{n_k-2}{2}$ and one understands how the fermionic represetation of 
table \ref{CFTTBA4} should be.\\
The $1/16$ sector has to be worked out expanding the Virasoro modes in integer fermionic modes, this is not very different from 
the previous situation, except for the presence of the zero-mode $\psi_0$ which generates the zero mode algebra:
\be \psi_0\big|1/16\big>_{-}=\frac{1}{\sqrt{2}}\big|1/16\big>_+\ee
\be \psi_0\big|1/16\big>_{+}=\frac{1}{\sqrt{2}}\big|1/16\big>_{-}\ee
So that actually there are 2 $1/16$ vacua, one has no 1-strings ($\big|1/16\big>_+$) and the other has 1 1-string 
($\big|1/16\big>_{-}$). The fermionic modes have indexes which are simply $\frac{n_k-1}{2}$, so that the $(1)$ quantum number is 
recognized as coming from the insertion of $\psi_0$.\\
 For example one has:
\be \big|1/16\big>_{-}= \sqrt{2}\psi_0\big|1/16\big>_+ \ee
\be L_{-1}\big|1/16\big>_{-}=\frac{1}{2\sqrt{2}}\psi_{-1}\big|1/16\big>_+\ee 
\be L_{-2}\big|1/16\big>_{-}=\frac{1}{\sqrt{2}}\psi_{-2}\big|1/16\big>_+\ee 
So that comparing with table \ref{h16c12} we easily understand the how things work out and, once again, aside from the 
normalizations one knows perfectly from the beginning which results he will find upon expressing eigenstates in terms of fermionic 
modes.\\
It has to be remarked that the fermionic description of the eigenstates of the RSOS transfer matrix fits naturally into the 
description of \cite{paths}, where it is shown that the TBA quantum numbers can be described also in terms of ``fermionic 
paths''.\\
Finally one notices that the TBA quantum numbers appearing in the $1/16$ and in the $1/2$ sector are actually the same, this can 
be understood from the existence of a boundary flow connecting the 2 sectors,  see for example \cite{lesage}.\\  

\section{Conclusion}
In this work we have derived the excited TBA equations for the $A_3$ model defined on a strip with integrable boundary conditions. The continuum limit of the TBA equations was then derived and an axpansion defined which led to integrals of motion. Such integrals of motion were identified to be the BLZ local integrals of motion. The identification was carried out by comparing the exact diagonalization of the Virasoro expressions with the spectrum derived form TBA. The eigenvalues were found to be exactly the same and the eigenvectors, once expressed in terms of fermionic modes, turned out to be labelled by the same quantum numbers as the eigenvalues.\\
It has to be remarked that this is the first time that an exact diagonalization of all the BLZ local integrals of motion has been carried out in a particular case including all the excited states.\\
A more detailed analisys of the spectra obtained would surely be interesting and will be the subject of successive work.

\section{Acknowledgments}
Alessandro Nigro would like to thank Sergio Caracciolo, Giuseppe Mussardo, Antonio Rago, Paolo Grinza for their encouragement and support. Acnkowledgement is due to Giovanni Feverati for his pointing the way into this beautiful field and for setting the $t=0$ initial conditions. Thanks are due as well to Paul Pearce and Changrim Ahn for useful discussions and advice, and to Bortolo Mognetti and Matteo Cardella for showing a couple of useful tricks.\\
Beyond professional acknowledgements, I would like to dedicate this work to my beloved Ruth Cominoli whose love i sacrificed for my pride, and lost in order to find the way into the unseen world which lies in science. 




\end{document}